\documentclass{article}
\usepackage{multido}
\usepackage{pgffor}

\usepackage{graphicx} % Required for inserting images
\usepackage{caption}
\usepackage{algorithm}
\usepackage{subcaption}
\usepackage{float}
\usepackage{mathtools}
\usepackage{amsmath}
\usepackage{algorithm}
\usepackage{algpseudocode}
\usepackage{tikz}
\usepackage{amssymb}
\newcommand{\vy}{\vec{{y}}}

\newcommand{\vr}{\vec{r}}
\def\pos{{\vec r}}
\def\psource{{\vec z}}

\newcommand{\vx}{\vec{\itbf{x}}}

\DeclareMathAlphabet{\itbf}{OML}{cmm}{b}{it}

\def\mC{\mathbb{C}}

\usepackage{mathrsfs}
\usepackage{algpseudocode}
\usepackage{bbm}
\usepackage{bm}
\usepackage{amsthm}
\theoremstyle{remark}

\usepackage{amsmath,amssymb,amsfonts}

\usepackage{tikz}
\usepackage{pgfplots}
\usepackage{url}
\usepgfplotslibrary{groupplots}
\usepackage{tikz}
\usetikzlibrary{positioning}
\usetikzlibrary{arrows.meta, bending, chains, decorations.markings}
\usepackage{comment}

\def\vect#1{\mbox{\boldmath{$#1$}}}

\def\vx{{\vec z}}
\def\psource{{\vec z}}

\def\vy{{\vec r}}
\def\pos{{\vec r}}
\newcommand{\bxx}{\vect x}
\newcommand{\byy}{\vect y}
\newcommand{\winc}{\psi^{i}}
\newcommand{\scatt}{\boldsymbol{\xi}}

\def\wg{{g}}

\newcommand{\wscatt}{\psi^{s}}
\newcommand{\wexc}{\psi^{e}}

\def\Gc{{\cal G}}

\newcommand{\C}{\mathbb{C}}

\algrenewcommand\algorithmicrequire{\textbf{Input:}}
\algrenewcommand\algorithmicensure{\textbf{Output:}}

\usepackage{siunitx}
\usepackage{hyperref}
\hypersetup{
    colorlinks=true,
    linkcolor=blue,
    filecolor=magenta,      
    urlcolor=cyan,
    }
\usepackage{xcolor}
\title{Imaging with super-resolution in changing random media}

\author{A. Christie, M. Leibovich, M. Moscoso, \\  A. Novikov, G. Papanicolaou and C. Tsogka}

\date{\today}

\begin{document}

\maketitle
\begin{abstract}
We develop an imaging algorithm that exploits strong scattering to achieve super-resolution in changing random media. The method processes large and diverse array datasets using sparse dictionary learning, clustering, and multidimensional scaling. Starting from random initializations, the algorithm reliably extracts the unknown medium properties necessary for accurate imaging using back-propagation, $\ell_2$ or $\ell_1$ methods. Remarkably, scattering enhances resolution beyond homogeneous medium limits. When abundant data are available, the algorithm allows the realization of super-resolution in imaging.
\end{abstract}
%\tableofcontents
%\sections{}
%\newpage

\bigskip

\begin{center} {\em Dedication}
\end{center}

\begin{quote}
  We dedicate this work to Professor Akira Ishimaru, whose  fundamental contributions have shaped the field of wave propagation in random media. 
\end{quote}

\section{Introduction}

High-resolution imaging from array data in unknown inhomogeneous ambient media requires estimating 
both the medium properties and the object characteristics. For diverse measurements collected from different sources in different, changing %random 
media, we introduce in this paper an algorithm that recovers the ambient %random 
media properties needed for high-resolution imaging as well as the source locations and strengths that constitute the imaging target. 

This algorithm extends and improves upon our previous work on imaging through random media using array data. Previously, we addressed imaging through a {\it single} unknown random medium, either weakly scattering \cite{Moscoso2024} or strongly scattering \cite{christie2026}. Here we consider the more challenging scenario of imaging through {\it multiple} strongly scattering random media, where super-resolution remains achievable as in the single-medium case \cite{christie2026}, while introducing essential algorithmic improvements.

Super-resolution refers to the phenomenon where the point spread function of an isolated source in a strongly scattering medium becomes significantly narrower than in a homogeneous medium using the same array imaging system, thereby providing enhanced resolution. While super-resolution in time-reversal has been extensively observed and analyzed \cite{Fink_2000,Borcea_2002}, time-reversal differs fundamentally from imaging. Time-reversal is a physical process: recorded array signals are time-reversed and physically re-emitted into the original medium, generating waves that back-propagate and refocus on the sources that produced them.

\begin{figure}[h]
    \centering
    \includegraphics[width=\linewidth]{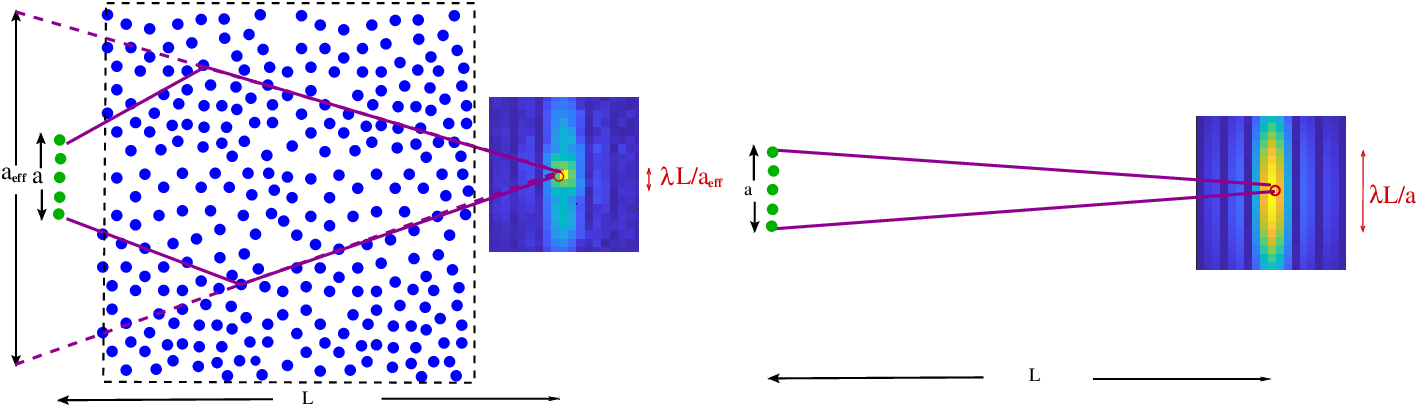}
    \caption{Schematic view of how super-resolution can be explained by the idea of a larger effective aperture \cite{Borcea_2002}. On the left the point spread function in the strongly scattering medium is illustrated and on the right the one obtained using the same array and bandwidth in the homogeneous medium.}
    \label{f:imagingschematic}
\end{figure}

Time-reversal does not constitute imaging precisely because neither the ambient medium nor the source locations are known, the physical refocusing occurs without forming an image. In strongly scattering media, time-reversal achieves tighter focusing than in homogeneous media (super-resolution) because scattering effectively enlarges the aperture, making the array data appear to originate from a larger, effective array \cite{Borcea_2002}, as illustrated schematically in Figure \ref{f:imagingschematic}.

In contrast, conventional imaging performs back-propagation analytically or numerically, typically assuming a homogeneous ambient medium despite the fact that the recorded data reflect wave propagation through an unknown scattering medium. To achieve super-resolution in imaging, one must therefore estimate the ambient medium properties. This is precisely what we achieved in \cite{christie2026} for a single random medium, developing an algorithm that exploits large, diverse array datasets.

The proposed imaging algorithm generalizes directly to the scattering case, that is, when small scattering objects are present in the imaging domain rather than sources. We can do this using the effective source formalism introduced in \cite{Moscoso14}.

The challenge in imaging when the ambient random medium is changing is that we do not assume that we know from which random medium the array data originates. We have developed an algorithm that is robust to how different the random media can be, that is, they can range from different, statistically independent random media to ones that are strongly correlated. This requires significant changes to the algorithm used for a single random medium in \cite{christie2026}.

\subsection{Array imaging with data from multiple random media}
The imaging problem in a single medium can be reduced to solving a linear system
$$ \vect{{\Gc}} \vect{x} = \vect{y}\, ,
$$
where $\vect{y}$ are the array data, an $N$-dimensional complex vector with $N$ the number of receivers in the array\footnote{For simplicity we describe here the single-frequency source imaging problem but in all computations we use multi-frequency data, so the array data for each frequency are stacked into a single data vector.}, $\vect{\Gc}$ is the $N\times K$-dimensional matrix of Green's functions with each column being the vector of Green's functions from a given source location to the array%in the %a single random medium
, and $\vect{x}$ is a complex $K$-dimensional vector that identifies the location and complex amplitude of the sources. The matrix of Green's functions $\vect{{\Gc}}$ is the sensing matrix. In a homogeneous medium the sensing matrix $\vect{\tilde{{\Gc}}}$ is known and has in three dimensions the form
$$
[\vect{{\tilde{\Gc}}}]_{j l} = \frac{e^{i k|\vy_j -\vx_l|}}{4\pi |\vy_j -\vx_l| }
$$
where $\{\vy_j\}_{j=1}^N$ are the coordinates of the receivers on the array, $\{\vx_l\}_{l=1}^K$ are the coordinates of the sources in the image window, and $k$ is the wavenumber.

In a random medium, both the sensing matrix $\vect{\Gc}$ and the vector $\vect{x}$ are to be determined, given the array data $\vect{y}$. This is an ill-posed problem, in general. Here we assume that a large and diverse collection of array data is available and the $N\times M$ data matrix is denoted by $\vect{Y}=[\vect{y}_1,\vect{y}_2,\ldots,\vect{y}_M]$, $M\gg K$. We also assume that the $K\times M$ source complex amplitude and location matrix  $\vect{X}=[\vect{x}_1,\vect{x}_2,\ldots,\vect{x}_M]$ is sparse, that is, each column $\vect{x}_i$ has small support less than or equal to an integer $s\ll K$. The matrix system to solve for a single %random 
random medium is then
$$ \vect{{\Gc}} \vect{X} = \vect{Y}\,,$$
with both $\vect{{\Gc}}$ and $\vect{X}$ to be determined. This is the sparse dictionary learning problem \cite{Engan1999,Rubinstein2010}, which is non-convex and, hence,  hard to solve because standard algorithms may get stuck in a sub-optimal solution.

When we have data from $L\geq 1$ different media with sensing matrices $\vect{\Gc}^{[i]}$, $i=1,2,\ldots,L$, and we do not know from what media the data come from, we denote the $\displaystyle N\times \sum_{i=1}^L M_i$ block data matrix by $\vect{Y}^B$,
%\textcolor{blue}{(it is not important but the size of this matrix is not necessarily N\times M\cdot L$, as we do not necessarily have $M$ data vetors from each $L$ different media)}, 
which is an unknown permutation of the block data matrix $[\vect{Y}^{[1]}, \vect{Y}^{[2]}, \dots, \vect{Y}^{[L]}]$ such that 
\begin{equation}
\label{eq:single}
 \vect{\Gc}^{[i]}\, \vect{X}^{[i]} = \vect{Y}^{[i]},\,~ i=1,2,\ldots,L.
\end{equation}
In block form the system to solve is now
\begin{equation}
\label{eq:block}
\vect{{\Gc}^B} \vect{X}^B = \vect{Y}^B\,,
\end{equation}
with $\vect{{\Gc}^B}$ an unknown permutation of the columns of  
\begin{equation}
\label{eq:blockG}
[\vect{\Gc}^{[1]}, \vect{\Gc}^{[2]}, \dots, \vect{\Gc}^{[L]}] \in C^{N\times K^B}
\end{equation}
and both $\vect{{\Gc}^B}$ and $\vect{X}^B \in C^{K^B\times M}$ are to be determined. Here $K^B=K\cdot L$ and we assume that the number of array data coming from the $i$-th medium, $M_i$, is large compared to the number of sources in the image window, $M_i \gg K$. We will assume for simplicity in the following that all $M_i$ are equal and denote by $M$ their sum over the $L$ media.

Within each single random medium sensing matrix $\vect{\Gc}^{[i]}$, the columns are relatively incoherent, that is, uncorrelated, reflecting the expected spatial arrangement of the sources in the image window as implied by the expected resolution of the imaging system \cite{borcea2003theory}.
Between different random sensing matrices $\vect{\Gc}^{[i]}$ and $\vect{\Gc}^{[j]}$, $i\neq j$, the columns could be arising from significantly correlated media to totally uncorrelated ones. They could, for example, come from slowly changing random media, which are correlated, or from fully independent random media. 

\subsection{The four-step imaging algorithm}

The imaging method has four steps. In the first step, we use a sparse dictionary learning algorithm  that yields accurate estimates of the columns of the block sensing matrix $\vect{\Gc}^B$, together with some extra noise columns\footnote{``Noise vectors or columns" in this context do not mean random noise in the traditional sense. Rather, they are linear combinations of true dictionary atoms that arise from the non-uniqueness of the optimization.}. This algorithm aims to find a sparse representation of the array data in the form of a linear combination of a few basic elements and it is described in Section \ref{ss:DL}. In array imaging in random media these basic elements are the Green's function vectors that form the columns of the block sensing matrix $\vect{\Gc}^B$. 

The second step uses clustering \cite{Ester1996} to distinguish between the estimates of the Green's function vectors and the extra noise vectors, so that we recover all the columns of $\vect{\Gc}^B$ accurately, as described in Section \ref{ss:greensclustering}. However, the columns of $\vect{\Gc}^B$ are unordered so they can not be used for imaging because (a) the random medium from which they come is not known, and (b) the points to which they correspond in the image window are also not known. 

In a third step, after the denoising by clustering, we separate the columns of $\vect{\Gc}^B$ into groups, each of which corresponds to a random medium $i$, $i=1,\dots, L$, using a new graph method as described in Section \ref{s:separation}. We then use non-metric multidimensional scaling \cite{Yi2003,Oh2010} in each random medium to associate the columns to their corresponding source points in the image window, thus ordering them. This methodology was introduced in imaging in \cite{Moscoso2024} and in \cite{christie2026}, and it is described in Section \ref{ss:mds}.

Once the block sensing matrix is determined and then ordered as described, it  can be used to produce $\ell_1$-norm \cite{chai2014imaging} or $\ell_2$-norm (Kirchhoff Migration) \cite{Borcea_2002} images. This is the fourth step. 
%The $\ell_1$-images are formed by associating the sparse coefficient values of each Green's function vector to the corresponding point in the imaging window. The $\ell_2$-images  are formed by backpropagating the data into the estimated media. 

The paper is organized as follows. In Section \ref{s:imagingprob}, we describe the imaging setup problem. In Section \ref{s:4stepalg}, we explain the four-step imaging method in detail. In Section \ref{s:numexp}, we present our numerical simulations.
Finally, Section \ref{s:conclusion} contains a summary and some conclusions.

\section{Formulation of the imaging problem}\label{s:imagingprob}
 In this section we discuss the model used in this paper to solve the forward wave propagation problem, and the generation of the data used. Suppose an array of $N$ receivers records signals emitted from sources located in a region of interest called the image window. Further suppose there is a collection of scatterers %in an otherwise uniform background 
 between the image window and the array, so the wavefront coming from the sources is distorted. 
 The array is taken to be one-dimensional in this paper. The direction parallel to the array is called the cross-range and the one perpendicular to the array is called the range. The propagation of a signal of frequency $\omega$ from point $\psource$ in the imaging window to a point $\vr$ is characterized by Green's function of the wave equation of the medium, $G(\pos, \psource)=G(\pos - \psource)$. For clarity of presentation, we omit its explicit dependence to the frequency $\omega$ in the paper.

\subsection{Data generation with the Foldy-Lax model}\label{ss:Foldy}

\begin{figure}[ht]
    \centering
    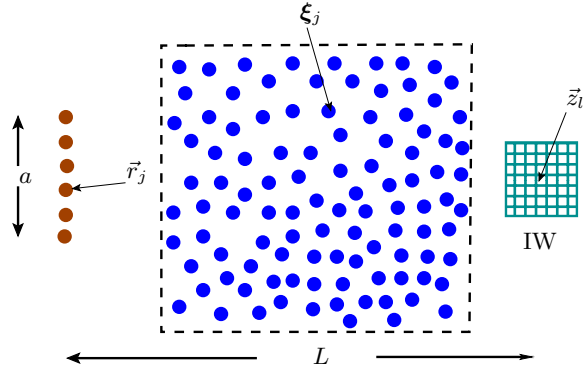
    \caption{Schematic of the imaging problem setup in a scattering medium. On the left, an array of receivers at locations $\vec{r}_j$, $j=1, \ldots,N$. The central square region contains randomly distributed scatterers at positions $\scatt_j$, $j=1,\ldots, J$. On the right, the imaging window (IW) is depicted, with discrete grid of points 
    $ \left\{ \vec{z}_l \right\}_{l=1}^K $  where the image will be reconstructed. }
    \label{fig:schematicFL}
\end{figure}

We use the Foldy-Lax equations~\cite{foldy,lax,Ishimaru1978Ch14,martin,let17} to model wave propagation in strongly scattering media. The Foldy-Lax model uses a discrete collection of randomly distributed scatters in an otherwise uniform background to simulate a multiple scattering medium (see Fig. \ref{fig:schematicFL}). %We now describe the Foldy-Lax equations. 

Define $\winc(\pos, \psource) = G_{0}(\pos - \psource)$ to be the incident wave at position $\pos$ from a source at position $\psource$, where
\begin{equation}
  G_{0}(\pos - \psource) : =  \frac{e^{i k |\pos - \psource|}}{4 \pi
    |\pos - \psource|}
\label{eq:incwave}
\end{equation}
is the free-space propagator and $k=\omega/c_0$ is the wavenumber with $c_0$ the constant wave speed. If there are
$J$ scatterers between the source at position $\psource$ and the receiver  at position $\pos$, then
the total wave $\psi(\pos, \psource)$ received at $\pos$ is the sum of the incident wave $\winc(\pos, \psource)$ and all the scattered waves $\wscatt_{j}(\pos)$ originated from 
the scatterer positions $\scatt_j$, $j=1,\dots,J$, so
\begin{equation}
  \psi(\pos, \psource)= 
  \winc(\pos, \psource)+ \sum_{j=1}^{J}
  \wscatt_j(\pos).
\label{eq:totalwave}
\end{equation}
%In equation,~\eqref{eq:totalwave} $\wscatt_{j}(\pos)$ is the wave that originates from a scatterer at position
%$\scatt_j$ and evaluated at position $\pos$. It is given by
Furthermore, the scattered wave $\wscatt_{j}(\pos)$ can be written as
\begin{equation}
  \wscatt_j(\pos) =
  G_{0}(\pos - \scatt_{j}) \tau_j \wexc_j,
  \label{eq:asymp}
\end{equation}
 where $\tau_{j}$ is the scattering
amplitude and $\psi_{j}^{e}$ is the  exciting wave at the scatterer position $\scatt_{j}$. Its dependence on the source location $\psource$ is not shown here. We also ignore self-interacting waves. Thus, the exciting wave at $\scatt_{j}$ is the sum of the scattered waves at all the other scatterers, and the incident wave $\winc(\scatt_{j}, \psource)$ at $\scatt_{j}$. This is written as
\begin{equation}
  \wexc_j =
  \winc(\scatt_j, \psource) +
  \sum_{m = 1,\ m \neq j}^{J} G_{0}(\scatt_{j} - \scatt_{m}) \tau_{m} 
  \wexc_{m}\,,
\label{eq:scattwave}
\end{equation}
provided that the scatterers are far enough apart from each other. 

For each fixed source location $\psource$, equation~\eqref{eq:scattwave} describes a consistent system of $J$ equations for the exciting waves $\wexc_1, \wexc_2, \dots \wexc_J$. The system can be solved using standard methods after rewriting it in matrix form.  Once, equation~\eqref{eq:scattwave} is solved we use equations~\eqref{eq:asymp} and  \eqref{eq:totalwave} to compute the scattered field at any position $\pos$. %In our numerical experiments we use $J=400$.
%We show the model set up in Figure~\ref{f:imagingschematic}.

We note that super-resolution is observed numerically in more complex and realistic multiple scattering models, such as in \cite{let17},  where finite size particles are considered rather than point scatterers, using a spherical harmonic expansion and a fast multipole algorithm for solving the resulting system of equations.

\subsection{Imaging with data from the Foldy-Lax model}\label{ss:setup}
Consider a single medium $i$, so  the scattering
amplitudes $\tau_{j}$ and  the scatterer locations $\scatt_{j}$ are fixed. Let
\begin{equation}\label{eq:GreenFuncVec}
\vect \wg^{(i)}(\vx)=[ \psi(\vy_{1},\vx),  \psi(\vy_{2},\vx),\ldots,
 \psi(\vy_{N},\vx)]^{T}\,
\end{equation}
be the Green's function vector that maps the data from a  source with unit complex amplitude located at $\vx$ to an array of $N$ receivers located at points $\vy_{1},\dots, \vy_{N}$. Even if we knew the source location and its complex amplitude, this vector is unknown because the scatterers are at unknown random locations. In equation~\eqref{eq:GreenFuncVec}, $\psi(\vy_{i},\vx)$ is the total field from a source at $\psource$ which is computed using the the Foldy-Lax model. 
The collection of Green's vectors for each source location $\vx_k$ forms the columns of the $N \times K$ sensing matrix of medium $i$, so
\begin{equation}\label{eq:sensingmatrix}
\vect{\Gc}^{(i)} =[\vect\wg^{(i)}(\vx_1)\,\cdots\,\vect\wg^{(i)}(\vx_K)] \ 
\end{equation}
is defined on a grid $\{\vx_k\}_{k=1}^K$  spanning the image window, with $K>N$, typically. The grid is unknown and, hence, it has to be determined from the array data.

The sensing matrix $\vect{\Gc}^{(i)}$
maps a distribution of sources in the image window to 
data received at the array in medium $i$. Indeed, if the vector $\bxx\in \mC^{K}$ has as $k$-th coordinate the complex amplitude of the source at location $\vx_k$ on the grid points, ${k=1,\dots,K}$, then
the data recorded on the array in the medium $i$ is given by
\begin{equation}
\label{eq:model}
\byy = \vect{\Gc}^{(i)}\, \bxx \, ,
\end{equation}
where $\byy \in \mC^{N}$ and  $\bxx$ is the image. In the case when multiple frequency data is recorded at the array, data for each
frequency is stacked into a single data vector. For simplicity, we describe the
single frequency case, but use multi-frequency data in the numerical simulations.

When the medium changes due to scatterer motion or other variations, data may be collected from $L$ different random configurations without knowledge of which configuration produced which measurements. To handle this uncertainty, we stack the Green's function vectors to form a block sensing matrix  $\vect{\Gc}^{B}$ as in \eqref{eq:blockG}, so that the array data can be written as
\begin{equation}\label{e:axeqb}
\byy = \vect{\Gc}^B\, \bxx \, ,
%~\vect{\Gc}^B=[\vect{\Gc}^{[1]}, \vect{\Gc}^{[2]}, \dots \vect{\Gc}^{[L]}],
\end{equation}
where %$\vect{\Gc}^{[i]}$ are the sensing matrices from the $L$ different medium realizations and 
\begin{equation}\label{e:Xi}    
 \bxx=
 \begin{bmatrix}
\bxx^{[1]}\\
\bxx^{[2]}\\
\vdots\\
\bxx^{[L]}\\
 \end{bmatrix}\,\,. %\, \bxx^{[i]}=\mathbf{0} \text{ for all but a single } i\in\{1, 2, \dots, L\}.
\end{equation}
Since each measurement originates from a single medium configuration, we have $\bxx^{[i]}=\mathbf{0} \text{ for all but one } i\in\{1, 2, \dots, L\}$.

\section{Analytical description of the imaging algorithm}\label{s:4stepalg}
To produce sharp images, we require accurate estimates of the columns of $\vect{\Gc}^{B}$ and their proper ordering.
We assume access to a large and diverse dataset comprising  $M\gg K\times L$ observations $\{\byy_m\}_{m=1,\dots,M}$, with $\byy_m = \vect{\Gc}^{B}\, \bxx_m$ as in (\ref{e:axeqb}). The source vectors $\bxx$'s are unknown, but assumed to be sparse.

In this section we explain the four steps of the imaging method. In the first step we use a dictionary learning algorithm to recover accurate estimates of the columns of the block sensing matrix $\vect{\Gc}^B$. The columns are recovered {along with extra noise column vectors in this step.} The second step separates the accurate column vectors from the noise vectors. Specifically, to refine the dictionary and eliminate the noise vectors, we repeat the first step multiple times with different random initializations and cluster the final results, removing the vectors that are not clustered. The first two steps, sparse dictionary learning plus clustering, recover the column vectors of the block sensing matrix accurately, but these are unordered, that is, they yield $\vect{\Gc^B}$ up to an unknown permutation of its columns.
%and a multiplication by complex scalars of unit length. 

In the third step, we cluster the recovered column vectors again into groups $\vect{\Gc}^{[i]}$ corresponding to the random media $i=1,\dots, L$. The clustering procedure is based on the recovered sparse coefficients in $\bm{X}$. 
%Since data points with intersecting $\ell_1$ supports must share a column vector, they must also come from the same random medium $i$. 
This is presented in section \ref{s:separation}. Finally, in the fourth step in section \ref{ss:mds}, we associate each column vector to a grid point in the image window using non-metric multidimensional scaling. This step is done separately for each, unordered, sensing matrix $\vect{\Gc}^{[i]}$ as in \cite{christie2026}. 

\subsection{Sparse dictionary learning }
\label{ss:DL}
An unordered estimate of the columns of the {block} sensing matrix $\vect{\Gc}^B\in\mathbb{C}^{N\times K^B}$ can be obtained, in principle, from the data matrix $\bm{Y}=[\bm{y}_1, \dots, \bm{y}_M]\in\mathbb{C}^{N\times M}$ by solving
the sparse dictionary learning problem 
\begin{equation}\label{e:modeq}
\min_{\bm{A}, \bm{X}} \|\bm{AX} - \bm{Y}\|_{F}^{2} \quad \text{s.t.} \quad \|\bm{x}_i\|_0 \leq s, \quad i = 1, \dots, M,
\end{equation}
for $\bm{A}\in\mathbb{C}^{N\times K^B}$ and $\bm{X}\in\mathbb{C}^{K^B \times M}$, provided $M\gg K^B$ is large enough and diverse.
Here, $\|\cdot\|_0$ is the number of nonzero elements, $s$ is the expected sparsity level, and $\|\cdot\|_F$ is the Frobenius norm. This problem is non-convex and is usually solved, after replacing the $\ell_0$-norm with its convex envelope, the $\ell_1$-norm, by alternating between an update for a sparse representation of the data $\bm{Y}$, given by the  coefficient matrix $\bm{X}$, and an update for the dictionary $\bm{A}$, so
\begin{equation}\label{e:sparsestep}
    \min\|\bm{X}\|_1 \text{ subject to }\bm{AX}=\bm{Y}
\end{equation}
and
\begin{equation}\label{e:l2step}
    \min_{\bm{A}}\|\bm{AX} - \bm{Y}\|_{F}^{2}
\end{equation}
are solved one after the other at each step of an iterative process. Problem \eqref{e:sparsestep} finds the sparsest representation for the data $\bm{Y}$ but requires the right dictionary $\vect{\Gc}^B$, which we do not know, and problem \eqref{e:l2step} finds the right dictionary assuming that the  coefficient matrix $\bm{X}$ is correct, which we also do not know. It is expected that by iterating between problems \eqref{e:sparsestep} and \eqref{e:l2step} the dictionaries $\bm{A}$ converge to the exact dictionary $\vect{\Gc}^B$, %one finds the right $\bm{A}$ and the right $\bm{X}$, 
up  to scaling and permutation.
%permutations of the columns of $\bm{A}$ and the corresponding rows of $\bm{X}$.

\subsubsection{Related work}\label{ss:rw}
The above problem has been extensively studied over the last decade, but only a few methods offer guarantees for {\em exact recovery}, and even these guarantees hold only under highly restrictive models or assumptions on the dictionary and the coefficient matrix. 
Spielman {\em et al.}~\cite{Spielman2013}, for example, established exact recovery when the dictionary matrix is square and nonsingular and the coefficient matrix has Bernoulli--Gaussian entries. 
Agarwal {\em et al.}~\cite{Agarwal2017} proposed an algorithm that guarantees approximate recovery of overcomplete, {\em incoherent} dictionaries. 
With additional postprocessing, arbitrarily accurate estimates of the dictionary can be obtained, provided the sparsity level is of order $O(N^{1/6})$ or lower for large $N$.
Novikov {\em et al.}~\cite{Novikov2023} also developed an algorithm with provable recovery guarantees under a suitable random model for both the dictionary and the coefficient matrices. 
Their method achieves exact recovery in polynomial time for sparsity levels that are linear in $N$ up to logarithmic factors, representing a significant improvement over previous approaches, including those cited above.

However, for general dictionaries, existing methods lack theoretical guaranties unless a good initialization is available—an important practical limitation. 
This limitation is also present in the Method of Optimal Directions (MOD)~\cite{Engan1999}, which converges to the exact solution only when preconditioned with an appropriate initial estimate of the dictionary columns~\cite{Agarwal2013}. 
Otherwise, the algorithm may converge to a suboptimal solution. 
The key feature of MOD, compared to other dictionary learning methods, is that at each iteration $n$, it updates the dictionary by solving a least-squares problem through the Moore--Penrose pseudoinverse of the current coefficient matrix $\bm{X}_n$, 
provided the corresponding matrices are invertible. 
MOD does not prescribe a specific method for solving the sparse coding step~\eqref{e:sparsestep}, which is typically handled by algorithms such as Matching Pursuit or Orthogonal Matching Pursuit. 
% Other sparse-promoting algorithms can in principle be used as well, but to the best of our knowledge, a detailed comparative study has not been published.

\subsubsection{Introduction of a modified MOD algorithm}\label{ss:ouralg}
The key differences between our sparse dictionary learning algorithm compared to other variants of MOD are: (1) we only do one iteration in the sparse optimization step \eqref{e:sparsestep}, so we do not let this algorithm converge, %for an approximate dictionary,  
(2) we use a significantly larger number of columns in $\bm{A}$ than the expected ones in $\vect{\Gc}^B$, and (3) we run the algorithm multiple times using different random initializations\footnote{{Regarding the initialization, numerical evidence suggests that the relaxation of the full $\ell_1$ minimization to a single step of $\ell_1$ minimization is the key to allowing for random initialization.}  } and (4) we do clustering at the end to remove noise columns and get a single vector for each ground truth column of $\vect{\Gc}^B$. With these new ingredients we always find the exact dictionary very accurately, and without the need of a preconditioning step so as to find a good initialization. 
%This is particularly important when the true dictionaries $\vect{G}^{[i]}$, are {\em coherent}, %i.e., when the inner products of their normalized columns are not small, 
%which is the case for imaging with high resolution because nearby grid points have normalized Green's function vectors whose inner products are close to one.

%{\color{green}{Alexei: I propose to remove "which is the case for imaging with high resolution..." because it is inaccurate and say: "which is the case for imaging in 
%time-varying media because Green's vectors remain coherent in time. " and then add}} 

More specifically, for a random initialization $\bm{A}_0\in\mathbb{C}^{N\times K_{MOD}}$ and $\bm{X}_0\in\mathbb{C}^{K_{MOD} \times M}$,  with $K_{MOD}=1.5\, K^B$, and
$\bm{Z}_0=\bm{0} \in\mathbb{C}^{N \times M}$ (see Algorithm {\bf 1}):
\begin{enumerate}
    \item We use one iteration of the Generalized Lagrangian Multiplier Algorithm (GeLMA)~\cite{Moscoso_2012}, so
\begin{eqnarray}
&&\bm{X}_{n+1}=\eta_{\tau \Delta t} \left( \bm{X}_{n} +\Delta t \bm{A}_n^{*}(\bm{Z}_{n}+\bm{Y}-\bm{A}_n\bm{X}_{n}) \right)
\, \label{fd2bis2} \\
&&\bm{Z}_{n+1} =  \bm{Z}_{n} + \Delta t  (\bm{Y} - \bm{A}_{n} \bm{X}_{n}) ,\nonumber 
\end{eqnarray}
where $\tau$ is the thresholding parameter, $\Delta t$ is the step size, and
\begin{equation}
\eta_a(x)=\begin{cases}
&x-a,~\hbox{ if } x>a,\\
&0,~\hbox{ if } -a< x<a,\\
&x+a,~\hbox{ if } x<-a\, \\
\end{cases}
\label{thresholding}
\end{equation}
is the shrinkage-thresholding operator which is applied component-wise. For a known $ \bm{A}_{n}$, GeLMA yields the exact solution for all values of the parameter $\tau$, independently of the data $\bm{Y}$.
    \item We set $\bm{A}_{n+1}=\bm{YX}_{n+1}^*(\bm{X}_{n+1}\bm{X}_{n+1}^*)^{-1}$, normalize the resulting columns to 1, and iterate until convergence.
\end{enumerate}

When the algorithm converges, the dictionary estimate $\bm{A}_{n+1}$ contains noise vectors in addition to the correct column vectors in the true dictionaries $\vect{G}^{[i]}$, $i=1,\dots,L$. 
%This is the case when the $\vect{G}^{[i]}}$ are coherent, so some of their columns are almost paralell. 
We remove the noise vectors by running the algorithm displayed below multiple times with different random initializations ($5$ in our numerical simulations). This provides multiple dictionary estimates, each containing accurate approximations of the columns of $\vect{\Gc}^B$, as well as a number of noise vectors which are removed 
%The noise vectors are removed 
using  clustering, as it is explained in the next subsection.

%\begin{figure}[H]
%\centering
\begin{minipage}{0.8\textwidth}
\begin{algorithm}[H]
%\caption{}
\begin{algorithmic}[1]
\State \textbf{Input}: Data matrix $\bm{Y} \in \mathbb{R}^{N \times M}$, step size $dt$, regulariztion parameter $\tau$, stopping criterion $\epsilon>0$, and an estimate of the number of columns $K_{MOD}$.
%Sparsity estimate $\hat{s}$, step size $dt$, thresholding parameter $\tau$, stopping criterion $\epsilon>0$, number of column estimates $K_{MOD}$, and data matrix $\bm{Y} \in \mathbb{R}^{N \times M}$ satisfying $\bm{Y} = \bm{A} \bm{X}$, where the columns of $\bm{X}$ are $s$-sparse.
\State \textbf{Output}: An unordered collection of columns $\{\hat{\bm{a}}_i\}_{i=1}^{{{K_{MOD}}}}$  containing estimates for the columns of matrix $\bm{A}$.
\State \textbf{Initialize}: ${\bm{X}_0}\in\mathbb{R}^{K_{MOD} \times M}$ uniformly at random, $\bm{Z}_0\in\C^{N \times M}$ with all entries equal to zero, ${\vect{A}}_0= \bm{Y} \hat{\bm{X}}_{0}^+$, normalize the columns to have length $1$, and set $n=0$.
%\State Normalize: The columns of $\hat{\vect{A}}_n$ to have length $1$.
\While{$|\bm{Y}- {\vect{A}}_n {\bm{X}}_n|>\epsilon$}
    \State Set: $\bm{Z}_{n+1}=(\bm{Y}-{\vect{A}}_n {\bm{X}}_n)dt+Z_n$
    \State Set: ${\bm{X}}_{n+1}  = {\vect{A}}_n^*(Y-{\vect{A}}_n {\bm{X}}_n+\bm{Z}_n)dt+{\bm{X}}_n$
    \State Set: ${\bm{X}}_{n+1} = \mathrm{sign}({\bm{X}}_{n+1})\eta(|{\bm{X}}_{n+1}|-\tau dt)$
    \State Set: ${\vect{\tilde{A}}}_{n+1}= \bm{Y} {\bm{X}}_{{n+1}}^+$
    \State Set: ${\vect{A}}_n$ to be the matrix ${\vect{\tilde{A}}}_{n+1}$ with columns normalized to length $1$.
    \State Set n=n+1
\EndWhile
\end{algorithmic}
\end{algorithm}
\end{minipage}
~\\
$\bm{X}^+$ denotes the pseudo-inverse of $\bm{X}$, $X^+ = X^*(\bm{XX}^*)^{-1}$.

\subsection{Improved estimate of sensing matrix by clustering}\label{ss:greensclustering}
An improved dictionary is constructed by clustering the columns of the multiple dictionaries found in the previous step using Density-Based Spatial Clustering of Applications with Noise (DBSCAN)~\cite{Ester1996} and selecting representatives from each cluster. Suppose that the set to be clustered is $\{\bm{a}_i\}_{i=1}^{\ell}$, where $\ell$ is the total number of normalized columns obtained from the previous dictionary learning step. Given a distance function $d$, threshold parameter $\varepsilon$, and minimal number of neighbors parameter $C_{min}$, the DBSCAN algorithm forms and expands clusters about areas with high point density and labels points in low-density areas as noise. If two points are within distance $\varepsilon$, then they are considered neighbors. If a point has $C_{min}$ neighbors, then it is considered a core sample in a cluster. The found columns are clustered as follows. 

Select a column $\bm{a}_i$ that has not been processed. If there are at least $C_{min}$ columns in the $\varepsilon$-ball $B_\varepsilon(\bm{a}_i)$, form a new cluster $C$ consisting of the columns in $B_\varepsilon(\bm{a}_i)$ that have not yet been clustered. If there are less that $C_{min}$ columns in the $\varepsilon$-ball, the column is labeled as noise. For each column $\tilde{\bm{a}}\in C$ with at least $C_{min}$ points in $B_\varepsilon(\tilde{\bm{a}})$ expand cluster $C$ by including the elements of $B_\varepsilon(\tilde{\bm{a}})$ that have not been assigned a cluster. The algorithm terminates after all data points have been processed. That is, after all columns are assigned to a cluster or assigned as noise.

The distance function used to cluster the normalized set of columns $\{\bm{a}_i\}_{i=1}^{\ell}$ is
\begin{equation}\label{e:dbsscandist}
d(\bm{a}, \tilde{\bm{a}})=1-|\left\langle \bm{a} , \tilde{\bm{a}}\right\rangle|. 
\end{equation}
Since the strenghts of the configurations are unknown, the recovered columns may differ by a phase factor. Using~\eqref{e:dbsscandist} as a metric corrects this issue since collinear vectors are considered the same. The final estimate of the unordered sensing matrix is obtained by taking a representative vector from each cluster. We use $C_{min}=5$ and $\epsilon=0.015$ for our numerical simulations. The number of clusters can be plotted against $\epsilon$ to determine what value leads to stable clustering. After this step we are left with $K^B$ %$\hat{\tilde{K}}$ 
unordered accurate estimates for the columns of $\vect{\Gc}^B$. In the next step we cluster this collection further into $L$ clusters corresponding to Green's function vectors from each individual random medium. %matrix $\vect{\Gc}^{[i]},~i=1, \dots, n$.
%associate the columns to focal points in the image window. %subdivide this collection into groups corresponding to columns in each individual sensing matrix $\vect{\Gc}^{[i]},~i=1, \dots, n$.

\subsection{Separating the medium realizations using graphs}\label{s:separation}
After the $K^B$ columns of $\vect{\Gc}^B\,$ are recovered we separate the columns into the $L$ sensing matrices $\vect{\Gc}^{[i]}$ by using the support of %the $M$ estimated sparse vectors in 
the coefficient matrix $\vect{X}^B \in C^{K^B\times M}$. Each column vector $\bxx_k$ of this matrix is an unknown permutation of the sparse vector
\begin{equation}
 \bxx_k=
 \begin{bmatrix}
\bxx^{[1]}\\
\bxx^{[2]}\\
\vdots\\
\bxx^{[L]}\\
 \end{bmatrix}\, ,
\end{equation}
where $\bxx^{[i]}$ is a vector that corresponds to an unknown source configuration in a medium $i\in\{1, 2, \dots, L\}$. Therefore, $\bxx^{[i]}=\mathbf{0}$ for all but a single $i\in\{1, 2, \dots, L\}$ and, thus, the $\bxx_k$'s with shared support 
\[\rho_k=\{j : \left |x_{k, j}\right |>0\}\]
come from the same medium $i$. Here, $x_{k, j}$ is the $j$-th component of vector $\bxx_k$. 
The reason why $\bxx^{[i]}=\mathbf{0}$ for all but a single $i\in\{1, 2, \dots, L\}$ is that each array data vector $\vect{y}_j$ arises from a linear combination of a few sources that are active in a single random medium.

To separate the columns of $\vect{\Gc}^B$ into their corresponding sensing matrices $\vect{\Gc}^{[i]}$ we construct a graph with nodes $\{1, 2, \dots, M\}$ corresponding to the data vectors $\vect{Y}=\vect{\Gc}^B \vect{X}^B$. The graph has an edge $\{i, j\}$  if there exists $\rho_{\ell}$ such that $i, j\in\rho_{\ell}$. The connected components of this graph correspond to distinct media. Indeed, if two Green's function vectors ${\bm{g}}_i$ and ${\bm{g}}_j$ are from different media, then there does not exist $\rho_\ell$ such that $i, j\in\rho_{\ell}$. Thus, ${\bm{g}}_i$ and ${\bm{g}}_j$ are in different connected components of the graph. 

After the Green's function vectors are separated into unordered sensing matrices $\hat{\vect{\Gc}}^{[i]}$ corresponding to each medium we reconstruct each grid separately. This is addressed in the next step.

\subsection{Grid reconstructions using multidimensional scaling}\label{ss:mds} 
We now describe the algorithm used to associate each estimated Green's function vector, the column vectors in the unordered sensing matrices $\hat{\vect{\Gc}}^{[i]}$, to grid points in the image window, so that we order the columns in each $\hat{\vect{\Gc}}^{[i]}$. 
This step is done separately for each medium $i$.  
%i.e., on each $\hat{\vect{\Gc}}^{[i]}$. Without this step, i.e., without ordering the columns of the sensing matrices, 
%imaging is not possible.

The Green's function vectors of each $\hat{\vect{\Gc}}^{[i]}$ can be associated with grid points in the image window using the connectivity based sensor localization algorithm in~\cite{Yi2003} and analyzed in~\cite{Oh2010}. This is a Multi-Dimensional Scaling (MDS) algorithm, with a proxy metric, used in~\cite{Moscoso2024} for imaging in a single weakly inhomogeneous random medium and in~\cite{christie2026} for imaging in a single scattering medium. To obtain sufficiently stable grid reconstructions in a strongly scattering medium the imaging system must operate with a suitable bandwidth and physical array size, as explained in ~\cite{christie2026} but not discussed further here.

The classical MDS algorithm takes as input the pairwise Euclidean distances between points (in our case the Green's function vectors) and returns the locations of these points in space, up to an overall rotation, translation and scaling. 
%a lower dimensional (in our case two-dimensional) representation of those points that best preserves the original distances. 
The classical MDS algorithm 
%the lower dimensional representation is obtained 
uses an eigenvalue decomposition~\cite{torgerson1952multidimensional}. 
%The lower dimensional representation 
The location of the points can also be found by minimizing the Frobenius norm between the input distance matrix and that of the pairwise distance of the points to be found using the SMACOF algorithm ("Scaling by MAjorizing a COmplicated Function")~\cite{Kruskal1964, Leeuw1977}. In this paper we use the SMACOF algorithm due to its better performance, but with a proxy metric as follows.

We use the geodesic graph distance of the edge-weighted fixed-radius nearest neighbors graph with radius $R$ as a distance function for MDS. The graph is constructed using the distance function defined in Eq.~\eqref{e:dbsscandist}. The cross-correlations in this distance function have the physical interpretation of time reversal experiments~\cite{Fink_2000}. If the signals recorded on an array are the (normalized) column vectors of the sensing matrices $\hat{\vect{\Gc}}^{[i]}$, and are re-emitted into the same medium they will focus at the location of the signal source due to the time-reversibility of the wave equation. Thus, computing the cross-correlation with the nearest neighbor is equivalent to using the nearest neighbor for time-reversal. Reconstructing the grid with cross-correlation connectivity information relies on this fundamental property of the wave equation and therefore is very robust to the inhomogeneities of the media.

More specifically, suppose that $\{\hat{\bm{g}}_i\}_{i=1}^{{K}}$ is the collection of estimated Green's vectors to be ordered, that is, associated with grid points in the image window. To reconstruct the grid, we first construct the connectivity graph. To this end, we fix a radius $R$ and define $\hat{\bm{g}}_i$ and $\hat{\bm{g}}_j$ to be neighbors if $d(\hat{\bm{g}}_i, \hat{\bm{g}}_i)<R$. Weight the edge between neighbors $\hat{\bm{g}}_i$ and $\hat{\bm{g}}_j$ with $d(\hat{\bm{g}}_i, \hat{\bm{g}}_j)$. The distance used in the MDS algorithm is the shortest path graph distance which can be found using, for example, Dijkstra's algorithm \cite{Dijkstra1959}. 

The radius $R$ should be selected based on the resolution of the imaging system. 
Through numerical testing, we found that the range $[0.6, .75]$ produces good results. We use $R=0.65$. We also found that specifying the number of neighbors performed worse than specifying the connectivity range.

After running MDS on each individual sensing matrix $\hat{\vect{\Gc}}^{[i]}$ we are left with a reconstructed grid of ${K}$ points for each realization of the medium $i$, $i=1,\dots,L$. Each reconstructed grid approximates the grid used to generate the data up to a translation, rotation, and scaling. To map the estimated grids onto a target grid we use Procrustes analysis~\cite{gower2004}. Specifically, we assume that we know the true location of $4$ anchor points. In numerical simulations we assume that these $4$ anchor points are the corners of the reconstructed grids. For each estimated grid we center and normalize the collection of estimated and known anchor points. Next, we map the standardized $4$ estimate anchor points onto the standardized $4$ known anchor locations using the linear transformation that minimizes the least squares distance between $4$ known anchor points. Finally, we map the full collection of points into the imaging window by performing the Procrustes transformation on the full collection of points.

\subsection{On the coherence of the columns of the block sensing matrix $\vect{\Gc}^B\,$}

A natural question concerns the coherence properties of the $K^B$ columns of $\vect{\Gc}^B\,$. For a single medium, we know that columns cannot be highly coherent for dictionary learning to succeed. However, multidimensional scaling requires some coherence to be able to identify neighbors. Thus, for each individual random medium, the sensing matrix $\vect{\Gc}^{[i]}$ must have columns that are relatively incoherent for sparse dictionary learning to work, yet retain sufficient coherence for MDS to work.

The coherence between columns of different media $\vect{\Gc}^{[i]}$ and $\vect{\Gc}^{[j]}$, for $i\neq j$ may range from large values for strongly correlated media to small values for completely independent configurations. 

This raises another question:  how can sparse dictionary learning succeed when dictionary elements are highly coherent? Our previous work on the role of vicinities in imaging ~\cite{MoscosoNovikovPapanicolaouTsogka2020} provides a theoretical framework that addresses this issue. In that work, we showed
that imaging can be performed effectively when scatterers are sufficiently separated for each measurement such that their vicinities do not overlap. The same analysis shows that sparse dictionary learning also succeeds under this condition. This requirement is naturally satisfied in changing media: a measurement cannot simultaneously contain signals from sources in two different media configurations. 

\section{Numerical simulations}\label{s:numexp}
The numerical simulations are conducted in the $C$-band radar regime using the Foldy-Lax model~\cite{foldy,lax,let17,martin}. Figure~\ref{fig:schematicFL} shows a schematic of the imaging system. The imaging system operates with a central frequency of $f_0=5$GHz, wavelength $\lambda=6$cm, bandwidth $B=1$GHz, and a frequency resolution of $df=40$MHz, resulting in $26$ frequencies. An array composed of $31$ transducers with distance $\lambda$ between them is placed $14$ meters from the center of the image window. The Green's function vector has dimensions $N=26\cdot 36 = 806$. The image window dimensions are defined in terms of the central wavelength. The cross-range spacing is $dl_x=0.5\lambda$ and the range spacing is $dl_z=1.66\lambda$. This produces a total of $K=361$ pixels ($nx=nz=19$).   The resolution in a homogeneous medium is estimated at $\lambda L/a\approx 8\lambda$ in cross-range and $5\lambda$ in range. On the other hand, the scattering area between the image window and the array is $10\times10$ meters and contains $J=400$ point scatterers.

\begin{figure}[htbp]
    \centering
    %\begin{subfigure}{0.9\textwidth}        
    \begin{subfigure}[t]{0.4\textwidth}
        \centering
        \includegraphics[width=.90\textwidth]{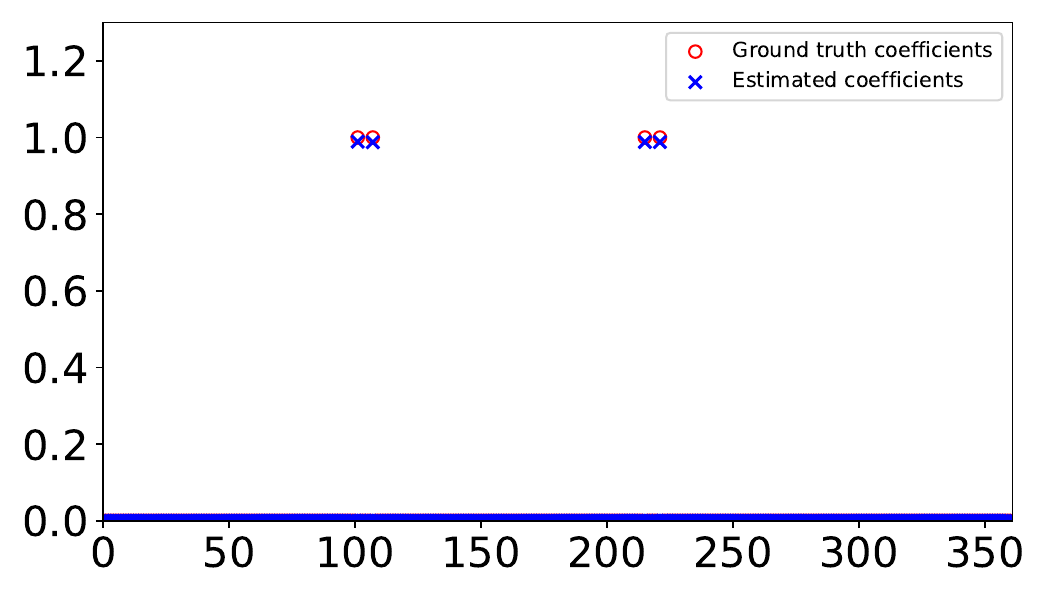}
    \end{subfigure}
%        \begin{subfigure}[t]{0.3\textwidth}
%        \centering
%        \includegraphics[width=.90\textwidth]{Figures/L1vL2/lambda05pert_n=3_l11d_diffmedia.pdf}
%    \end{subfigure}
    \begin{subfigure}[t]{0.4\textwidth}
        \centering
        \includegraphics[width=.90\textwidth]{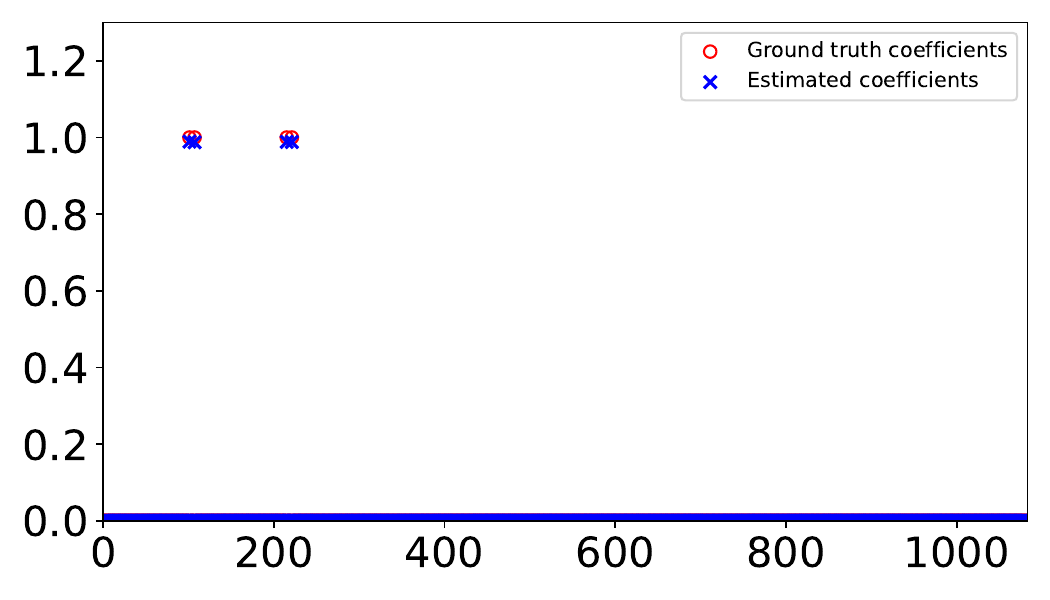}
    \end{subfigure}
        \begin{subfigure}[t]{0.4\textwidth}
        \centering
        \includegraphics[width=.90\textwidth]{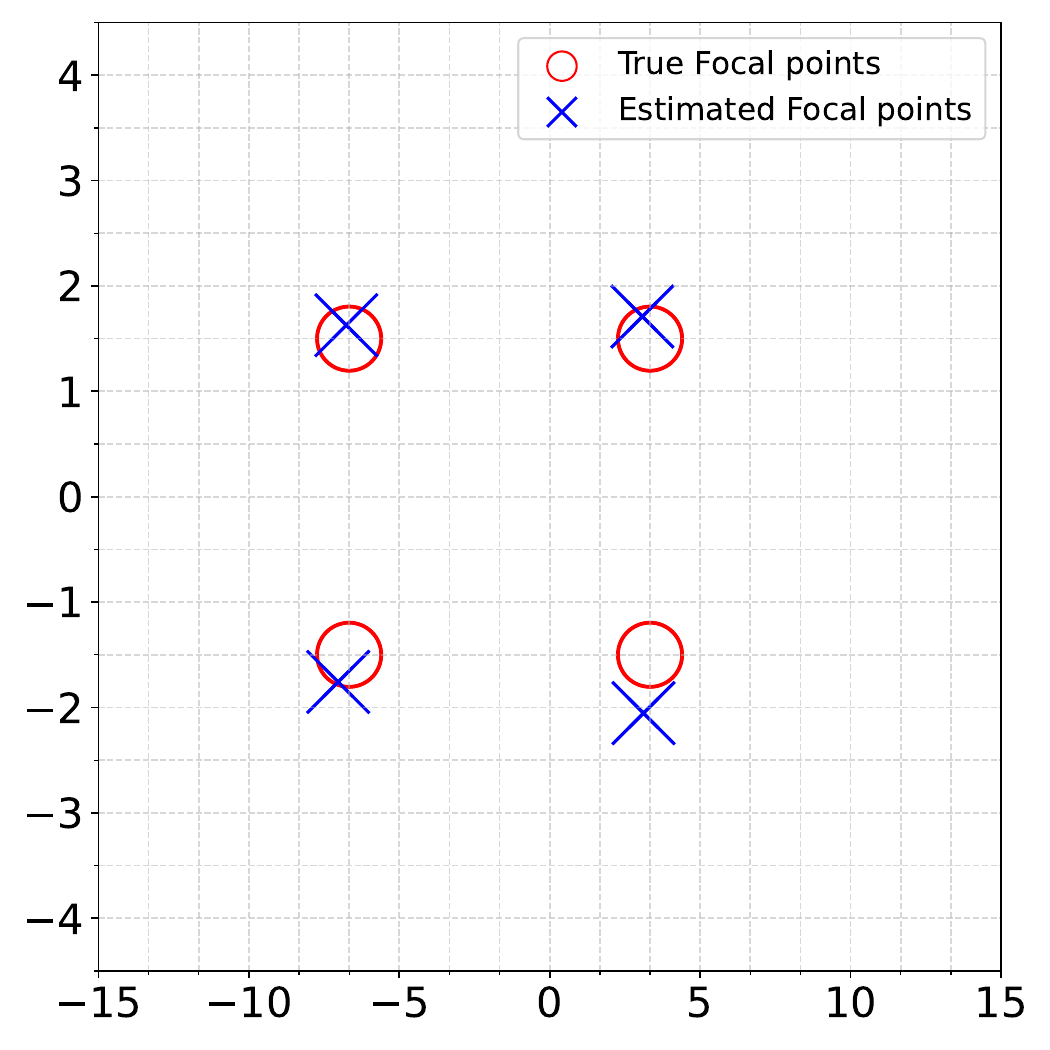}
    \end{subfigure}
%        \begin{subfigure}[t]{0.3\textwidth}
%        \centering
%        \includegraphics[width=.90\textwidth]{Figures/L1vL2/lambda05pert_n=3_l1top4_diffmedia.pdf}
%    \end{subfigure}
    \begin{subfigure}[t]{0.4\textwidth}
        \centering
        \includegraphics[width=.90\textwidth]{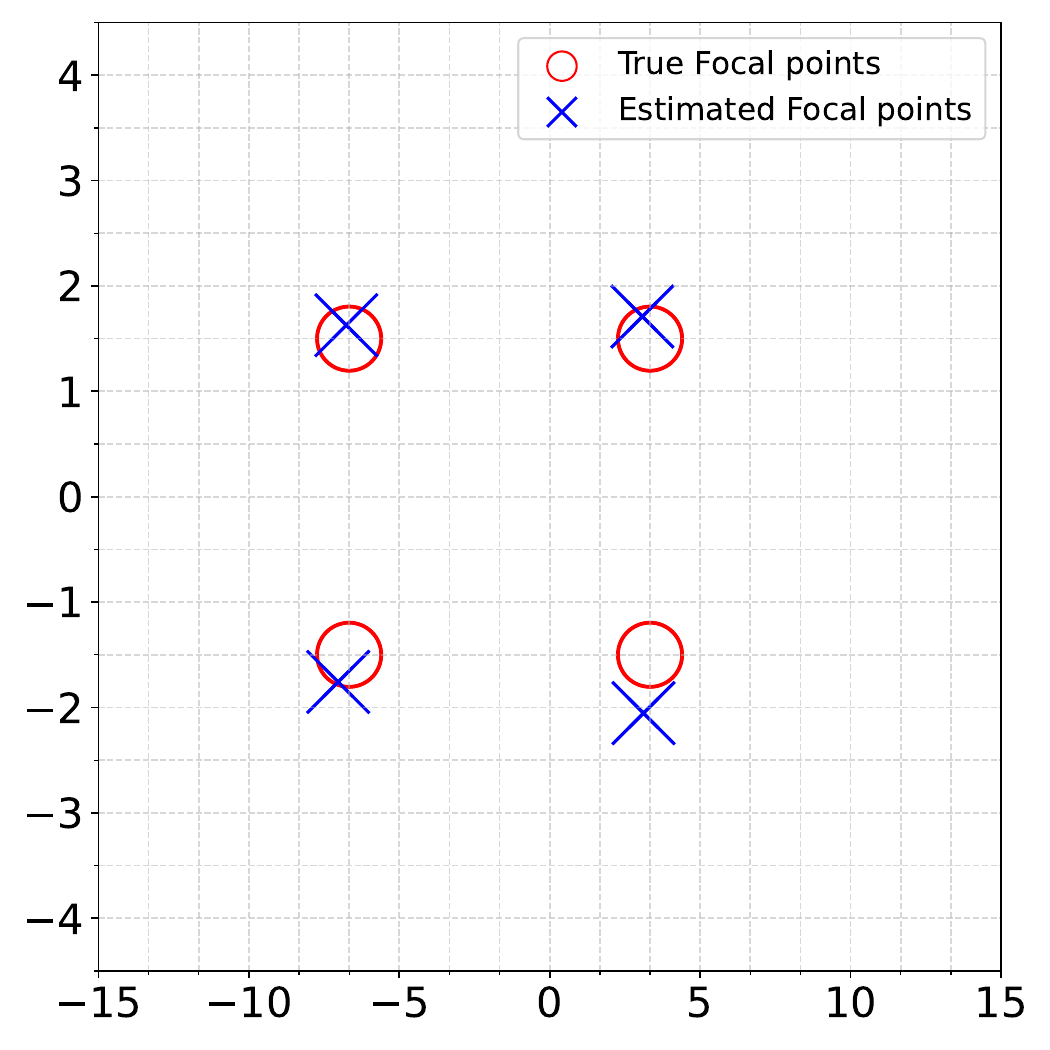}
\end{subfigure}
\begin{subfigure}[t]{0.4\textwidth}
        \centering
        \includegraphics[width=.90\textwidth]{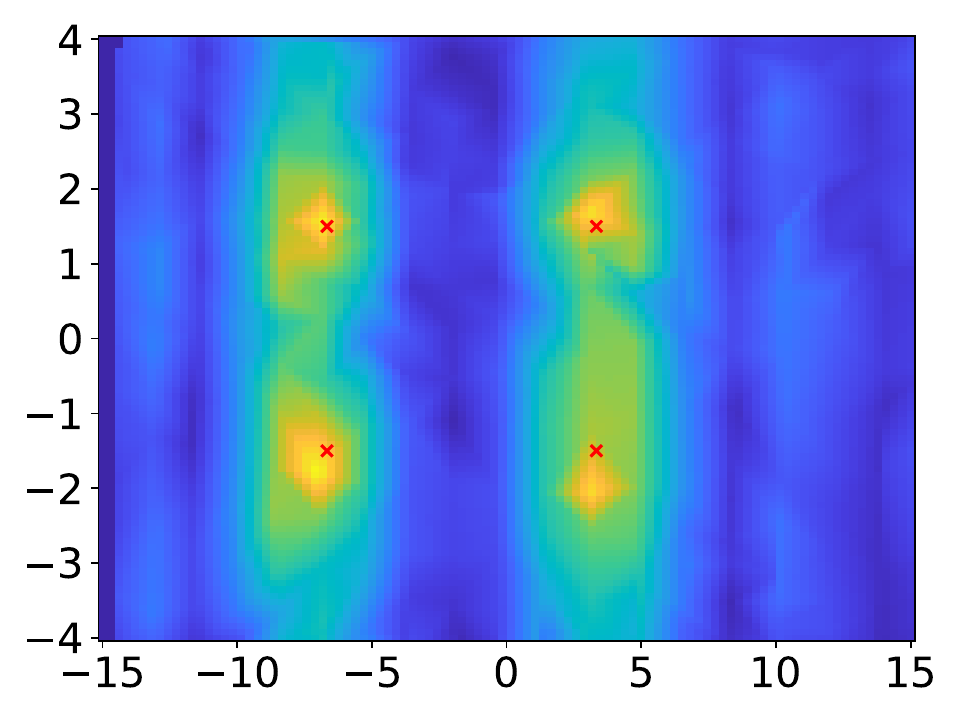}
    \end{subfigure}
%    \begin{subfigure}[t]{0.3\textwidth}
%        \centering
%        \includegraphics[width=.90\textwidth]{Figures/L1vL2/lambda05pert_KM_n=3_diffmedia.pdf}
%    \end{subfigure}
    \begin{subfigure}[t]{0.4\textwidth}
        \centering
        \includegraphics[width=.90\textwidth]{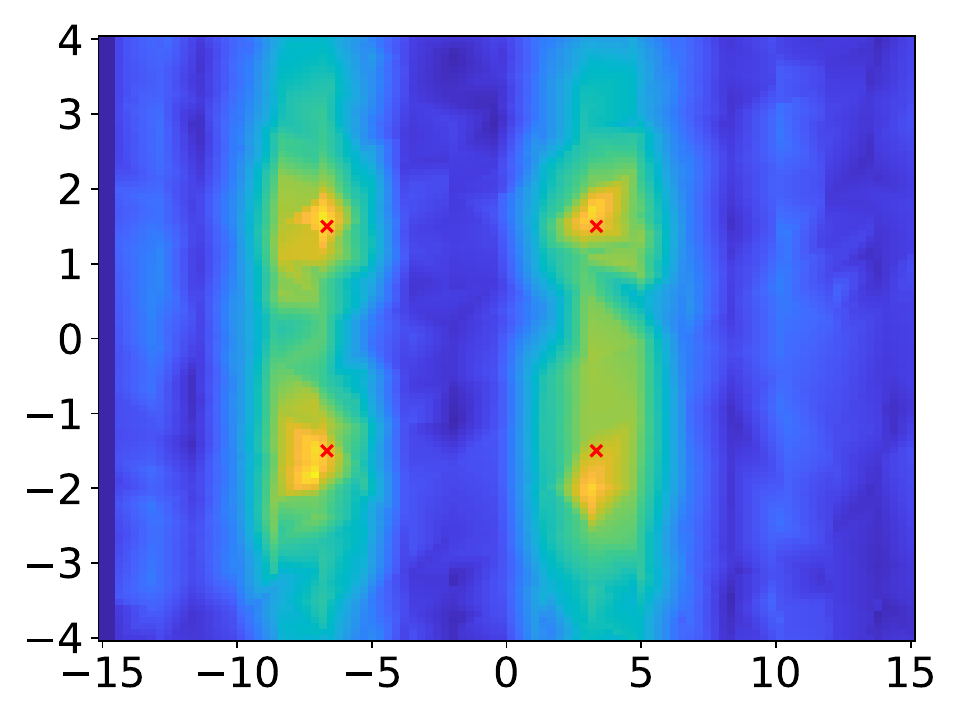}
    \end{subfigure}
    \caption{$\ell_1$- and KM-images for $L=3$ \textbf{nearby (correlated) media.} 
     \textbf{Top row}: the solution $\bm{x}^{\ell_1}$ obtained with $\ell_1$ minimization . 
    \textbf{Center row}: Locations of top $4$ $\ell_1$-norm coefficients plotted with blue $x$'s and the true sources locations plotted with red circles. \textbf{Bottom row}: KM-images. \textbf{Left column}: Images produced using the actual medium from which the data originated. 
    %\textbf{Center column}: Images produced using an independent random medium %with scatterers placed uniformly at random in the scattering region. 
    \textbf{Right column}: Images produced using the combined sensing matrix $\vect{\hat{\Gc}}^B$. In the KM images the true sources locations are indicated with red x's}
        \label{f:l1l2nearby}
\end{figure}

The data received at the array are from $s$ sparse source configurations in the imaging window, so no more than $s$ sources are simultaneously active in the imaging window. Each source is taken to have a random complex amplitude with modulus $1$ and their locations are randomly sampled in the grid of the imaging window. We assume that there is an abundance of data from each grid point. In numerical experiments we take $s=4$, have on average $10$ measurements for each grid point, and have $M=3610$ measurements for each random medium. This results in $10LK=10830$ measurements in total.

\begin{figure}[htbp]
    \centering
    %\begin{subfigure}{0.9\textwidth}        
    \begin{subfigure}[t]{0.4\textwidth}
        \centering
        \includegraphics[width=.90\textwidth]{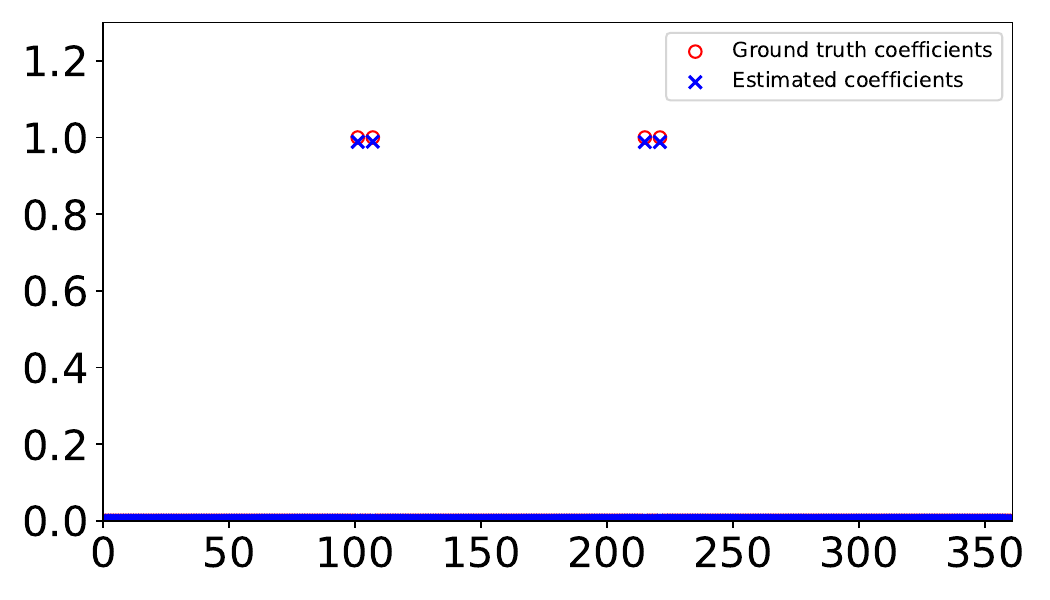}
    \end{subfigure}
%        \begin{subfigure}[t]{0.3\textwidth}
%        \centering
%        \includegraphics[width=.90\textwidth]{Figures/L1vL2/Uniformscatters_n=3_l11d_diffmedia.pdf}
%    \end{subfigure}
    \begin{subfigure}[t]{0.4\textwidth}
        \centering
        \includegraphics[width=.90\textwidth]{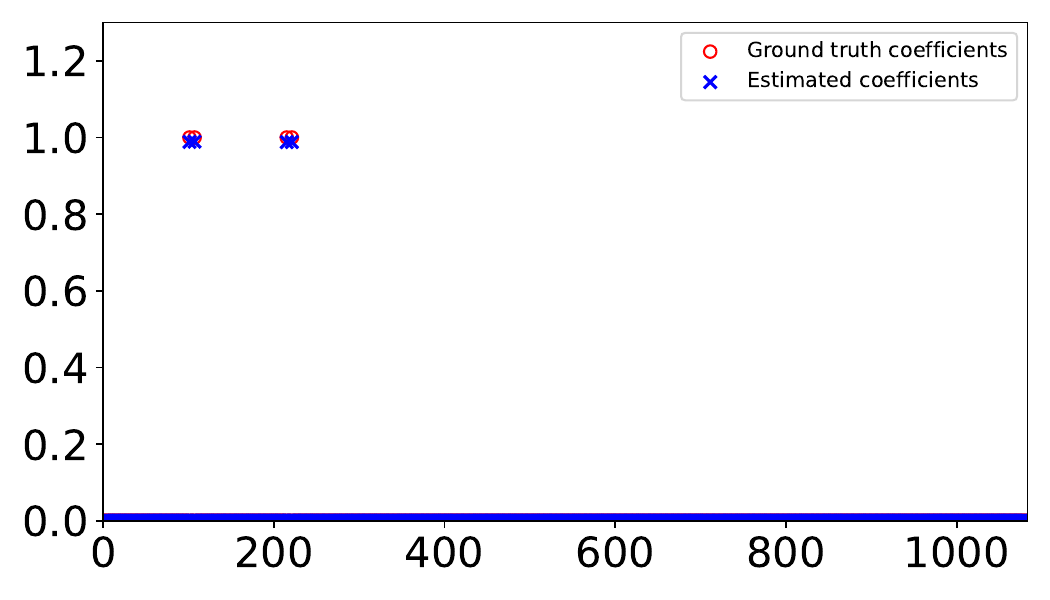}
    \end{subfigure}
  \begin{subfigure}[t]{0.4\textwidth}
        \centering
        \includegraphics[width=.90\textwidth]{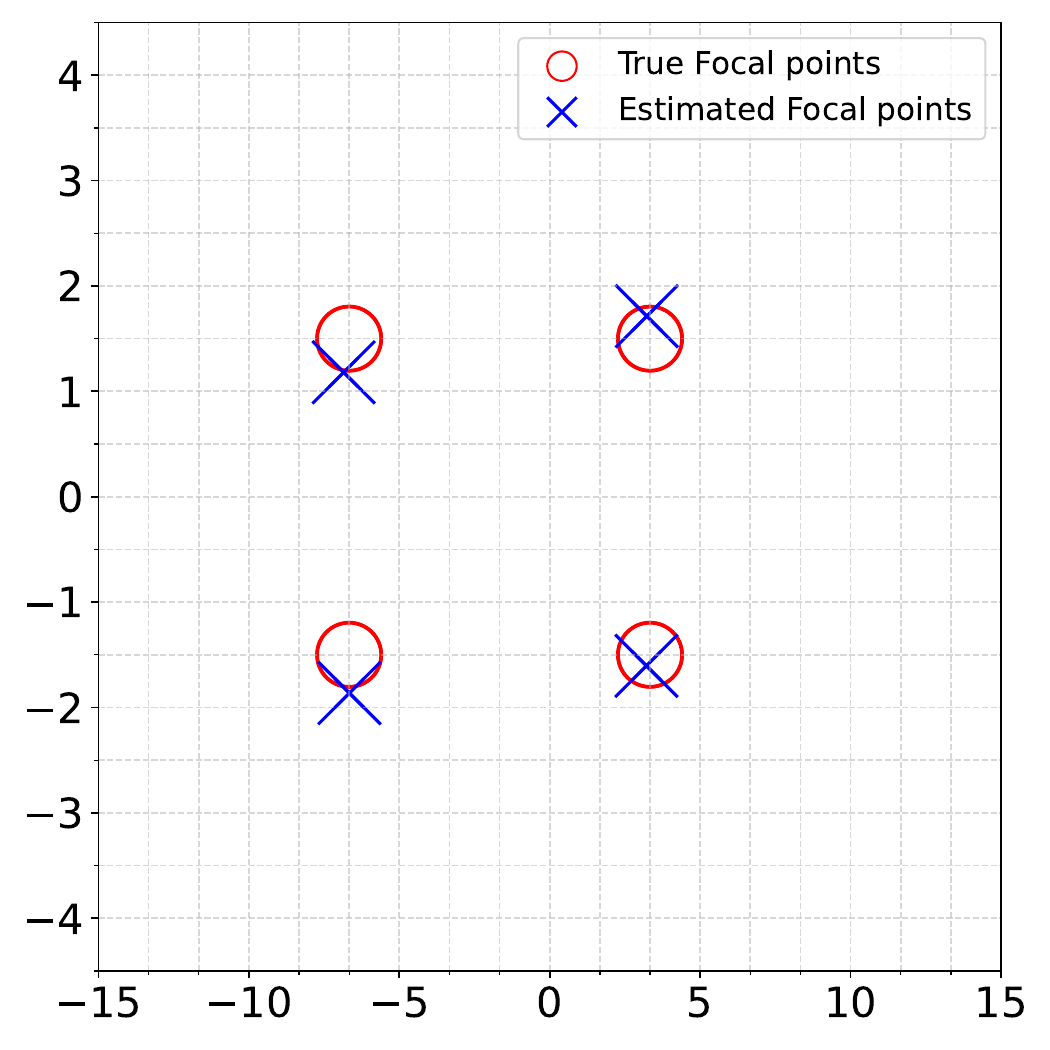}
    \end{subfigure}
%        \begin{subfigure}[t]{0.3\textwidth}
%        \centering
%        \includegraphics[width=.90\textwidth]{Figures/L1vL2/Uniformscatters_n=3_l1top4_diffmedia.pdf}
%    \end{subfigure}
    \begin{subfigure}[t]{0.4\textwidth}
        \centering
        \includegraphics[width=.90\textwidth]{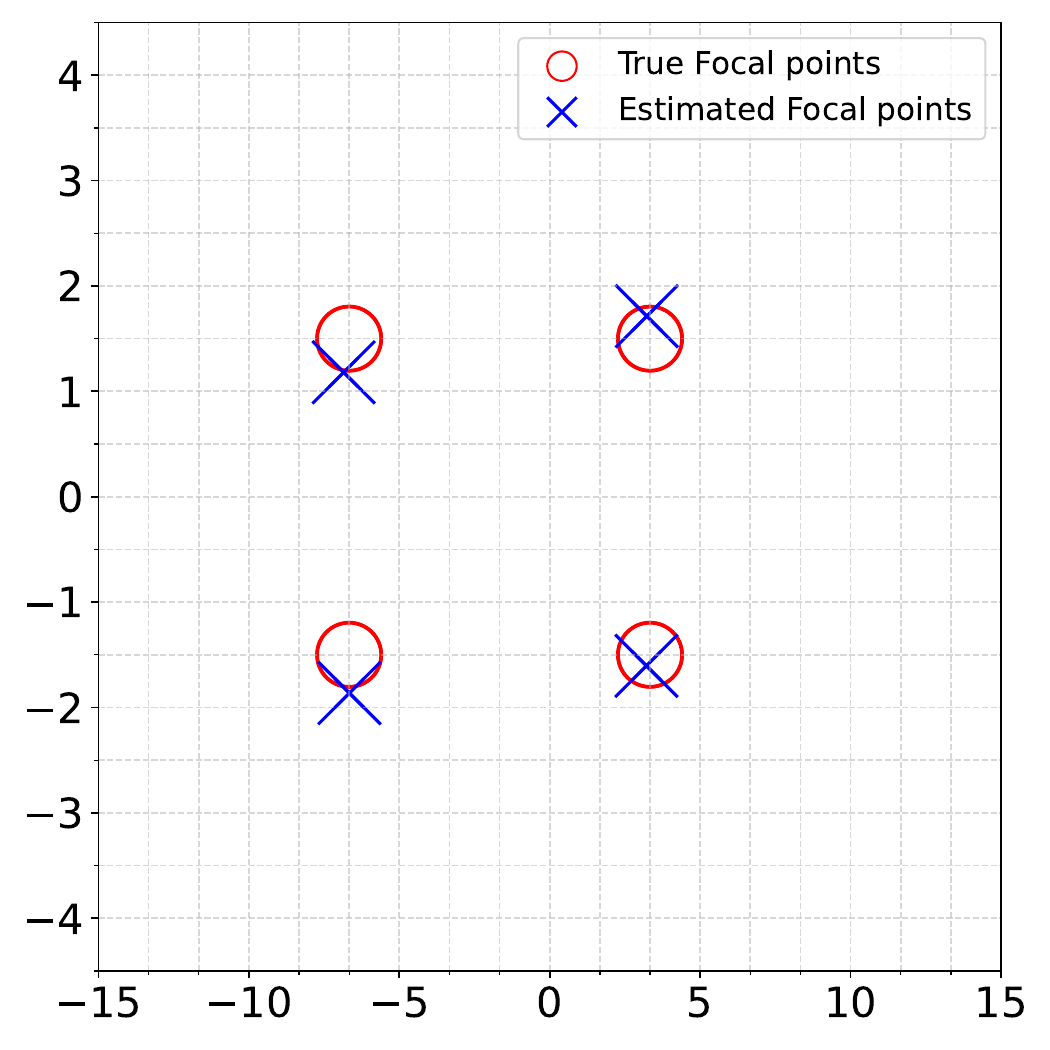}
    \end{subfigure}    
    \begin{subfigure}[t]{0.4\textwidth}
        \centering
        \includegraphics[width=.90\textwidth]{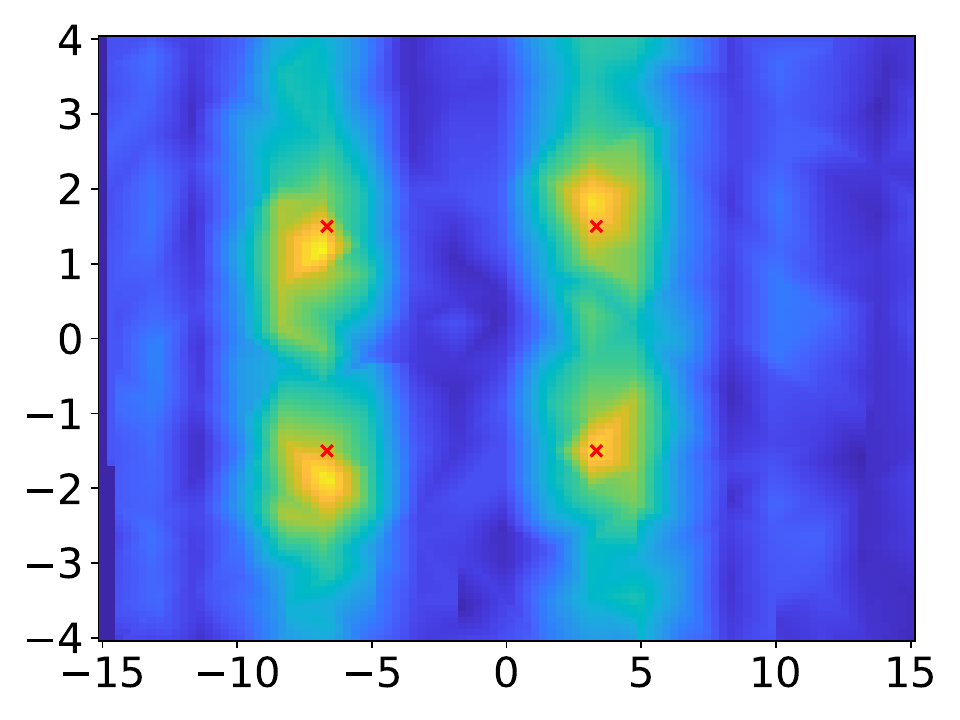}
    \end{subfigure}
%    \begin{subfigure}[t]{0.3\textwidth}
%        \centering
%        \includegraphics[width=.90\textwidth]{Figures/L1vL2/Uniformscatters_KM_n=3_diffmedia.pdf}
%    \end{subfigure}
    \begin{subfigure}[t]{0.4\textwidth}
        \centering
        \includegraphics[width=.90\textwidth]{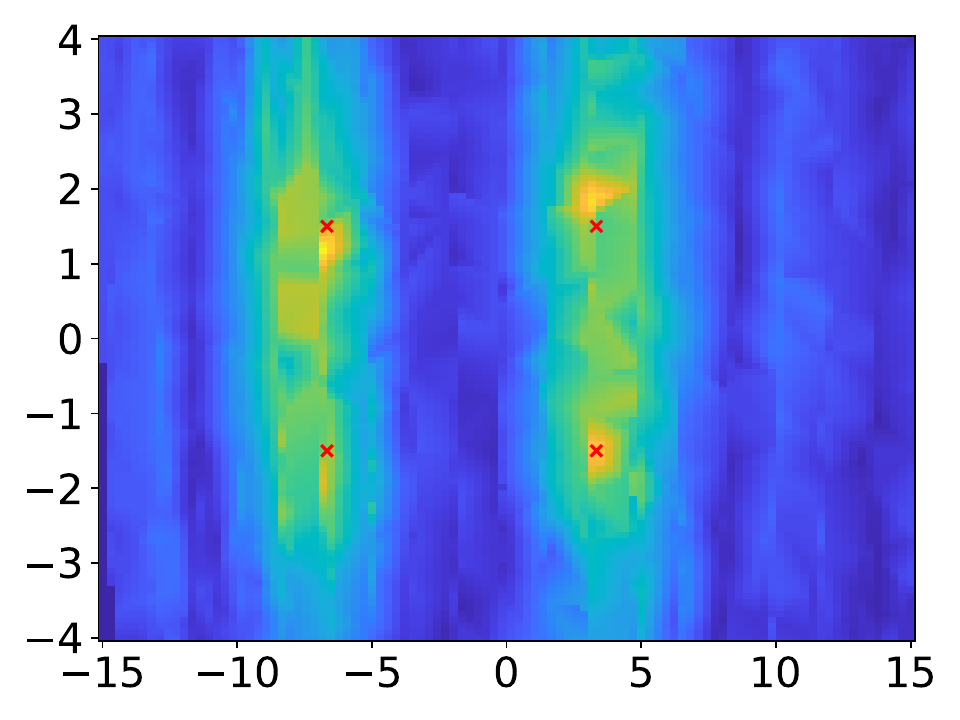}
    \end{subfigure}

    \caption{$\ell_1$ and KM images for $L=3$ {\bf independent media.} \textbf{Top row}: the solution $\bm{x}^{\ell_1}$ obtained with $\ell_1$ minimization. 
    \textbf{Center row}: Locations of top $4$ $\ell_1$-norm coefficients plotted with blue $x$'s and the true sources locations plotted with red circles. \textbf{Bottom row}: KM-images. \textbf{Left column}: Images produced using the actual medium from which the data originated. 
    %\textbf{Center column}: Images produced using an independent random medium %with scatterers placed uniformly at random in the scattering region. 
    \textbf{Right column}: Images produced using the combined sensing matrix $\vect{\hat{\Gc}}^B$. In the KM images the true sources locations are indicated with red x's}
    \label{f:l1l2independent}
\end{figure}

In Figures~\ref{f:l1l2nearby} and \ref{f:l1l2independent} we consider two different situations for imaging in changing random media depending on the correlation between them. 
In Figure~\ref{f:l1l2nearby} we consider three nearby or correlated media, whose scatterer positions are small random perturbations of the scatterer positions of the other media. The positions of the scatterers $\scatt_j^{[i]}$ of media $i=2,3$  are perturbed by adding random vectors drawn from $N(0, (0.05\lambda)^2)$ to the scatterer positions $\scatt_j^{[1]}$ of medium $i=1$. 
In Figure~\ref{f:l1l2independent}, however, the three media are independent or uncorrelated, which means that their scatterers are placed uniformly at random in the scattering region.

After grouping the Green's function vectors by medium and reconstructing the grid for each, we form images by  
combining the ordered sensing matrices $\vect{\hat{\Gc}}^{[i]}$ into a block matrix 
$$\vect{\hat{\Gc}}^B = \left[\vect{\hat{\Gc}}^{[1]}, \vect{\hat{\Gc}}^{[2]}, \dots, \vect{\hat{\Gc}}^{[L]}\right]$$ and solving the large system 
$$\vect{\hat{\Gc}}^B \bm{x} = \bm{y}.$$

Images can be formed either by using $\ell_1$-minimization (top and middle rows in Figs.~\ref{f:l1l2nearby} and~\ref{f:l1l2independent}) or $\ell_2$-minimization (bottom rows in Figs.~\ref{f:l1l2nearby} and~\ref{f:l1l2independent}). The $\ell_1$-minimization images are formed by solving the corresponding linear problems using GeLMA, so we associate the sparse coefficient values of each Green's vector to the corresponding grid point in the imaging window. The  $\ell_2$-minimization images are formed using  Kirchhoff Migration (KM). 
For comparison, the left columns of Figures~\ref{f:l1l2nearby} and~\ref{f:l1l2independent} show results obtained when we know which medium produced the measurements. This is the ideal single-medium case where the ambient medium is not changing.  
%Applying $\big(\vect{\hat{\Gc}}^{B}\big)^{*}$ to the data $\bm y$ yields $\bm x^{\mathrm{KM}}\in\mathbb{C}^{K\times L}$. We write $\bm x^{\mathrm{KM}}=[\bm x^{\mathrm{KM},1},\ldots,\bm x^{\mathrm{KM},L}]$, where each $\bm x^{\mathrm{KM},i}\in\mathbb{C}^{K}$ is the KM image computed on its corresponding grid. The KM images shown in the bottom-right panels of Figs.~\ref{f:l1l2nearby} and \ref{f:l1l2independent} are displayed after linear interpolation (Python/SciPy’s \texttt{griddata}) onto a composite grid formed by concatenating the $L$ individual grids (total $KL$ points). The bottom-left panels are obtained by applying $\big(\vect{\hat{\Gc}}^{i}\big)^{*}$ to the data $\bm y$ where $i$ denotes the true medium that generated the data; interpolation is performed on that medium’s grid. These serve as the ideal KM images obtained when we know from which medium the data originated. 
Applying the conjugate transpose of $\vect{\hat{\Gc}}^{B}$ to the data $\bm{y}$ produces a vector $\bm{x}^{KM}$ of dimension $K \times L$.  This corresponds to $L$  KM images, $\bm{x}^{KM,i}$, $i=1,\ldots,L$ each one constructed on the corresponding grid. The KM images shown in the bottom-right plots of Figs.~\ref{f:l1l2nearby} and~\ref{f:l1l2independent} are produced using linear interpolation (Python/SciPy’s \texttt{griddata}).  The grid used for interpolation is the size $L K$  grid obtained from concatenating the $L$ grids from each medium. %The KM images in the bottom-left plots of Figs.~\ref{f:l1l2nearby} and~\ref{f:l1l2independent} are produced using the dictionary corresponding to the actual medium from which the data originated and the corresponding grid is used for interpolation. These are the ideal KM images we could obtain if we knew from which medium the data originated.  
%In the right column of Figs.~\ref{f:l1l2nearby} and~\ref{f:l1l2independent} we display at each point $\vec{z}_l$ the value $\displaystyle \frac{1}{L} \sqrt{\sum_{i=1}^{L}  | \bm{x}^{KM,i}(\vec{z}_l) |^2} $} \textcolor{magenta}{Alex can you ceck if this is what you do ? That is what Alexei said. To be honest I still don't undertsand what is plotted since we do not have one grid but $L$ versions of it. So to me it is not clear what $KM$ is. Can you clarify please ?  

When the media are correlated, see Figure~\ref{f:l1l2nearby}, both $\ell_1$ minimization and KM produce excellent results comparable to the single medium case. For independent media, however, see Figure~\ref{f:l1l2independent},  the $\ell_1$-minimization image remains excellent because the correct Green's function vectors are used to select the appropriate grid points, while incorrect vectors do not interfere with the minimization. The support recovery of the $\ell_1$ image is exact. In contrast, KM imaging degrades substantially (bottom-right image), producing less useful results than the single-medium case  (bottom -left).

%\textcolor{magenta}{Thoughts: Do we really need to show the middle columns in Fig 2 and 3? I think we should only propose strategy 2 for image reconstruction because when new data $y$ come in we do not know from which medium they are so 1. does not make sense. We can show the left column in these figures because this is the "ideal" result we would get if we new from which medium the data is from. Why does KM in Fig 3 bottom- right does not work ? not sure I understand ?} 
%
%
%#####################
\section{Summary and conclusions}\label{s:conclusion}

In this paper, we introduce an algorithm to form high-resolution images 
%from array data when the ambient medium is randomly inhomogeneous and 
when a large and diverse collection of data from different sources in different, changing random media is available.
The algorithm uses sparse dictionary learning and clustering to recover an unordered collection of estimates of Green's function vectors and
a multidimensional scaling algorithm that associates them to grid points in the image window. The key is a variant of the MOD algorithm for dictionary learning that, when followed with the DBSCAN clustering algorithm, provides very accurate estimates of the Green's function vectors even when these are highly coherent. This is essential for imaging with high resolution. In addition, this algorithm  does not require a good initialization, and converges for any initial random guess of the dictionary and the coefficient matrix.

After these two steps, we are left with highly accurate Green's  function vectors that are clustered again into groups based on the media from which they come. The collections of Green's function vectors from each media are then associated to grid points in the image window using non-metric multidimensional scaling, producing multiple reconstructed grids. The main idea of the grid reconstruction step is to use cross-correlation information between the Green's  function vectors to construct a connectivity graph. The grids are then reconstructed using multidimensional scaling on the shortest path distance of the graph, similar to what is done in sensor network localization problems. 

The association between Green’s-function vectors and image-window grid points enables the ordering of the sensing matrices for each random medium, so they can be used to form $\ell_1$- and KM-images. We find that the $\ell_1$ reconstruction with the combined sensing matrix $\vect{\Gc}^B$ yields the best imaging results. The accurate Green's vector estimates allow accurate estimation of the sparse image configuration. Furthermore, this approach does not require knowledge of what medium the data came from.

An important extension of this work arises when large, diverse training sets are unavailable for each medium. In that case, we restrict attention to slowly varying media so that information learned from one medium can be used for imaging in a nearby medium. These extensions are currently under investigation.

%The association between Green's function vectors and grid points in the image window means that the sensing matrices of each random media are now ordered, so these can be used to produce $\ell_1$- or KM-images.  
%The $\ell_1$-images are constructed by associating the sparse coefficient values of each Green's vector to the corresponding grid point in the imaging window. The KM-images are constructed by backpropagating the data using the estimated Green's funtions, interpolating the pixel values between grid points and then selecting the best image. 
%We find that using the combined sensing matrix $\vect{\Gc}^B$ to produce $\ell_1$-images gives the best results. The highly accurate Green's vector estimates allow accurate estimation of the sparse image configuration. Furthermore, $\ell_1$-images constructed from the combined sensing matrix do not require knowledge of what medium the data came from.

%An important extension of this work is when we do not have large and diverse data sets for each random medium that is encountered. In this case we need to restrict to media that are changing slowly and then some information from one medium can be used for imaging in a nearby medium. These extensions are currently under investigation.

\bigskip
{\bf Acknowledgements.} Miguel Moscoso's work was supported by  the Spanish AEI grant PID2020-115088RB-I00.
Alexei Novikov's work was partially supported by NSF 2407046, AFOSR FA9550-23-1-0352, and FA9550-23-1-0523. 
The work of Alexander Christie and George Papanicolaou was partially supported by AFOSR FA9550-23-1-0352.
The work of  Chrysoula Tsogka was partially supported by AFOSR FA9550-23-1-0352 and FA9550-24-1-0191.

% bibliography

\bibliographystyle{ieeetr}
\bibliography{bib.bib}

@article{christie2026,
author = {Alexander Christie  and Matan Leibovich  and Miguel Moscoso  and Alexei Novikov  and George Papanicolaou  and Chrysoula Tsogka },
title = {Data-driven superresolution imaging in disordered media},
journal = {Proceedings of the National Academy of Sciences},
volume = {123},
number = {1},
pages = {e2530449123},
year = {2026},
doi = {10.1073/pnas.2530449123},
URL = {https://www.pnas.org/doi/abs/10.1073/pnas.2530449123},
eprint = {https://www.pnas.org/doi/pdf/10.1073/pnas.2530449123}}

@inbook{Ishimaru1978Ch14,
  author    = {Akira Ishimaru},
  title     = {Multiple Scattering Theory of Waves in Stationary and Moving Scatterers and Its Relationship with Transport Theory},
  booktitle = {Wave Propagation and Scattering in Random Media},
  publisher = {Academic Press},
  year      = {1978},
  volume    = {2},
  chapter   = {14},
  address   = {New York}
}

@INPROCEEDINGS{MOD,
author={Engan, K. and Aase, S.O. and Hakon Husoy, J.},
booktitle={1999 IEEE International Conference on Acoustics, Speech, and Signal Processing. Proceedings. ICASSP99 (Cat. No.99CH36258)}, 
title={Method of optimal directions for frame design}, 
year={1999},
volume={5},
number={},
pages={2443-2446 vol.5},
doi={10.1109/ICASSP.1999.760624}}

@inproceedings{Ester1996,
author = {Ester, Martin and Kriegel, Hans-Peter and Sander, J\"{o}rg and Xu, Xiaowei},
title = {A density-based algorithm for discovering clusters in large spatial databases with noise},
year = {1996},
publisher = {AAAI Press},
booktitle = {Proceedings of the Second International Conference on Knowledge Discovery and Data Mining},
pages = {226–231},
numpages = {6},
location = {Portland, Oregon},
series = {KDD'96}
}

@INPROCEEDINGS{Engan1999,
  author={Engan, K. and Aase, S.O. and Hakon Husoy, J.},
  booktitle={1999 IEEE International Conference on Acoustics, Speech, and Signal Processing. Proceedings. ICASSP99 (Cat. No.99CH36258)}, 
  title={Method of optimal directions for frame design}, 
  year={1999},
  volume={5},
  number={},
  pages={2443-2446 vol.5},
  doi={10.1109/ICASSP.1999.760624}}

@incollection{MoscosoNovikovPapanicolaouTsogka2020,
  author       = {Miguel Moscoso and Alexei Novikov and George Papanicolaou and Chrysoula Tsogka},
  title        = {Data Structures for Robust Multifrequency Imaging},
  booktitle    = {First Congress of Greek Mathematicians: Proceedings of the Congress held in Athens, Greece, June 25–30, 2018},
  editor       = {Ioannis Emmanouil and Anargyros Fellouris and Apostolos Giannopoulos and Sofia Lambropoulou},
  pages        = {181--230},
  year         = {2020},
  publisher    = {De Gruyter},
  address      = {Berlin ; Boston},
  doi          = {10.1515/9783110663075-009}
}

@article{Dijkstra1959,
  title={{A note on two problems in connexion with graphs}},
  author={Dijkstra, E.W},
  journal={Numer. Math.},
  volume={1},
  number={},
  pages={269–271},
  year={1959}
}

@ARTICLE{Rubinstein2010,
       author = {{Rubinstein}, R. and {Bruckstein}, A. M. and {Elad}, M.},
        title = "{Dictionaries for Sparse Representation Modeling}",
      journal = {Proceedings of the IEEE},
         year = 2010,
        month = {june},
       volume = {98},
       number = {6},
          eid = {},
        pages = {1045-1057},
          doi = {10.1109/JPROC.2010.2040551},
       adsurl = {https://ui.adsabs.harvard.edu/abs/2011InvPr..27h5004B}
}

@article{borcea2003theory,
  title={Theory and applications of time reversal and interferometric imaging},
  author={Borcea, Liliana and Papanicolaou, George and Tsogka, Chrysoula},
  journal={Inverse Problems},
  volume={19},
  number={6},
  pages={S139},
  year={2003},
  publisher={IOP Publishing}
}

@INPROCEEDINGS{Oh2010,
  author={Oh, Sewoong and Montanari, Andrea and Karbasi, Amin},
  booktitle={2010 IEEE Information Theory Workshop on Information Theory (ITW 2010, Cairo)}, 
  title={Sensor network localization from local connectivity: Performance analysis for the MDS-MAP algorithm}, 
  year={2010},
  volume={},
  number={},
  pages={1-5},
  doi={10.1109/ITWKSPS.2010.5503144}}

@inproceedings{Yi2003,
author = {Shang, Yi and Ruml, Wheeler and Zhang, Ying and Fromherz, Markus P. J.},
title = {Localization from mere connectivity},
year = {2003},
isbn = {1581136846},
publisher = {Association for Computing Machinery},
address = {New York, NY, USA},
url = {https://doi.org/10.1145/778415.778439},
doi = {10.1145/778415.778439},
booktitle = {Proceedings of the 4th ACM International Symposium on Mobile Ad Hoc Networking \& Computing},
pages = {201–212},
numpages = {12},
location = {Annapolis, Maryland, USA},
series = {MobiHoc '03}
}

@article{Moscoso_2012,
doi = {10.1088/0266-5611/28/10/105001},
url = {https://dx.doi.org/10.1088/0266-5611/28/10/105001},
year = {2012},
month = {9},
publisher = {IOP Publishing},
volume = {28},
number = {10},
pages = {105001},
author = {Miguel Moscoso and Alexei Novikov and George Papanicolaou and Lenya Ryzhik},
title = {A differential equations approach to l1-minimization with applications to array imaging},
journal = {Inverse Problems}
}

@article{chai2014imaging,
  title={Imaging strong localized scatterers with sparsity promoting optimization},
  author={Chai, Anwei and Moscoso, Miguel and Papanicolaou, George},
  journal={SIAM Journal on Imaging Sciences},
  volume={7},
  number={2},
  pages={1358--1387},
  year={2014},
  publisher={SIAM}
}

@ARTICLE{Agarwal2017,
  author={Agarwal, Alekh and Anandkumar, Animashree and Netrapalli, Praneeth},
  journal={IEEE Transactions on Information Theory}, 
  title={A Clustering Approach to Learning Sparsely Used Overcomplete Dictionaries}, 
  year={2017},
  volume={63},
  number={1},
  pages={575-592},
  doi={10.1109/TIT.2016.2614684}}

@book{gower2004,
    author = {Gower, John C and Dijksterhuis, Garmt B},
    title = {Procrustes Problems},
    publisher = {Oxford University Press},
    year = {2004},
    month = {01},
    isbn = {9780198510581},
    doi = {10.1093/acprof:oso/9780198510581.001.0001},
    url = {https://doi.org/10.1093/acprof:oso/9780198510581.001.0001},
}

@article{Fink_2000,
doi = {10.1088/0034-4885/63/12/202},
url = {https://dx.doi.org/10.1088/0034-4885/63/12/202},
year = {2000},
month = {12},
publisher = {},
volume = {63},
number = {12},
pages = {1933},
author = {Mathias Fink and Didier Cassereau and Arnaud Derode and Claire Prada and Philippe Roux and Mickael Tanter and Jean-Louis Thomas and François Wu},
title = {Time-reversed acoustics},
journal = {Reports on Progress in Physics}
}

@article{Kruskal1964,
  author       = {J. B. Kruskal},
  title        = {Multidimensional scaling by optimizing goodness of fit to a nonmetric hypothesis},
  journal      = {Psychometrika},
  year         = {1964},
  volume       = {29},
  number       = {1},
  pages        = {1--27},
  doi          = {10.1007/BF02289565},
  url          = {https://doi.org/10.1007/BF02289565},
  issn         = {1860-0980}
}

@article{Leeuw1977,
author = {de Leeuw, Jan},
year = {1977},
month = {01},
pages = {},
title = {Application of Convex Analysis to Multidimensional Scaling},
journal = {Recent Developments in Statistics}
}

@article{Moscoso2024,
author = {Miguel Moscoso  and Alexei Novikov  and George Papanicolaou  and Chrysoula Tsogka },
title = {Correlation-informed ordered dictionary learning for imaging in complex media},
journal = {Proceedings of the National Academy of Sciences},
volume = {121},
number = {11},
pages = {e2314697121},
year = {2024},
doi = {10.1073/pnas.2314697121},
URL = {https://www.pnas.org/doi/abs/10.1073/pnas.2314697121},
eprint = {https://www.pnas.org/doi/pdf/10.1073/pnas.2314697121}}

@article{torgerson1952multidimensional,
  title={Multidimensional scaling: I. Theory and method},
  author={Torgerson, William S.},
  journal={Psychometrika},
  volume={17},
  number={4},
  pages={401--419},
  year={1952},
  publisher={Springer}
}

@article{foldy,
  author = {L. L. Foldy},
  title = {The multiple scattering of waves},
  journal = {Phys. Rev.},
  volume = {67},
  pages = {107--119},
  year = {1945}
}

@article{lax,
  author = {M. Lax},
  title = {Multiple scattering of waves},
  journal = {Rev. Mod. Phys.},
  volume = {23},
  pages = {287--310},
  year = {1951}
}

@article{let17,
  author = {P.-D. Letourneau and Y. Wu and G. Papanicolaou and J. Garnier and E. Darve},
  title = {A numerical study of super-resolution through fast 3D wideband algorithm for scattering in highly-heterogeneous media},
  journal = {Wave Motion},
  volume = {70},
  pages = {113--134},
  year = {2017}
}

@book{martin,
  author = {P. A. Martin},
  title = {Multiple Scattering: Interaction of Time-Harmonic Waves with N Obstacles},
  publisher = {Cambridge University Press},
  year = {2006}
}

@article{Borcea_2002,
doi = {10.1088/0266-5611/18/5/303},
url = {https://dx.doi.org/10.1088/0266-5611/18/5/303},
year = {2002},
month = {8},
publisher = {},
volume = {18},
number = {5},
pages = {1247},
author = {Liliana Borcea and George Papanicolaou and Chrysoula Tsogka and James Berryman},
title = {Imaging and time reversal in random media},
journal = {Inverse Problems}
}

@inproceedings{Spielman2013,
author = {Spielman, Daniel A. and Wang, Huan and Wright, John},
title = {Exact recovery of sparsely-used dictionaries},
year = {2013},
isbn = {9781577356332},
publisher = {AAAI Press},
booktitle = {Proceedings of the Twenty-Third International Joint Conference on Artificial Intelligence},
pages = {3087–3090},
numpages = {4},
location = {Beijing, China},
series = {IJCAI '13}
}

@article{Agarwal2013,
  title={Exact Recovery of Sparsely Used Overcomplete Dictionaries},
  author={Alekh Agarwal and Anima Anandkumar and Praneeth Netrapalli},
  journal={ArXiv},
  year={2013},
  volume={abs/1309.1952},
  url={https://api.semanticscholar.org/CorpusID:9275907}
}

@article{Novikov2023,
title = {Dictionary Learning for the Almost-Linear Sparsity Regime},
journal = {Proceedings of Machine Learning Research vol 201:1–39, 2023},
volume = {201},
number={},
pages = {1-39},
year = {2023},
issn = {1063-5203},
doi = {},
url = {},
author = {Alexei Novikov and Stephen White}
}

@article{Moscoso14,
  author    = {A. Chai and M. Moscoso and G. Papanicolaou},
  title     = {Imaging strong localized scatterers with sparsity promoting optimization},
  journal   = {SIAM Journal on Imaging Sciences},
  volume    = {7},
  pages     = {1358--1387},
  year      = {2014}
}

%\newpage
%\printbibliography
\end{document}